\documentclass[prd,twocolumn,showpacs,amsmath,amssymb]{revtex4}

\usepackage{graphicx}
\usepackage{dcolumn}
\usepackage{bm}
\def\Journal#1#2#3#4{{#1} {\bf #2}, #3 (#4)}
\def\PRD{Phys. Rev. D}
\def\PRL{Phys. Rev. Lett.}
\def\NPB{Nucl. Phys. B}
\def\PLB{Phys. Lett. B}
\def\APJ{Astrophys. J.}
\def\APJS{Astrophys. J. S}
\def\APP{Astropart. Phys.}
\def\MPLA{Mod. Phys. Lett. A}
\def\PRep{Phys. Rep.}
\def\etal{{\sl et al.}}

\begin{document}

\title{Accessibility of the Pre-Big Bang Models to LIGO}

\author{Vuk Mandic}
\affiliation{LIGO Laboratory, California Institute of Technology, MS 18-34, Pasadena, CA 91125}

\author{Alessandra Buonanno$^{a,b}$}

\affiliation{$^a$ Department of Physics, University of Maryland, College Park, MD 20742 \\  
$^b$ AstroParticule et Cosmologie (APC)~\footnote{UMR 7164 (CNRS, Universit\'e Paris 7, CEA, 
Observatoire de Paris)} 11, place Marcelin Berthelot, 75005 Paris, France}

\date{\today}

\begin{abstract}
The recent search for a stochastic background of gravitational waves
with  LIGO interferometers has produced a new upper bound on the
amplitude of this background in the 100 Hz region. We investigate
the implications of the current and future LIGO results on pre-Big-Bang 
models of the
early Universe, determining the exclusion regions in the parameter space of 
the minimal pre-Big Bang scenario. Although the current LIGO reach is still 
weaker than the 
indirect bound from Big-Bang nucleosynthesis, future runs by LIGO, in 
the coming year, and by 
Advanced LIGO ($\sim$2009) should further constrain the parameter space, 
and in some parts surpass the Big-Bang nucleosynthesis bound. 
It will be more difficult to constrain the parameter space in non-minimal 
pre-Big-Bang models, which are characterized by multiple cosmological phases 
in the yet not well understood stringy phase, and where the higher-order 
curvature and/or quantum-loop corrections in the string effective action should be included.
\end{abstract}


\maketitle

\section{Introduction}

The Laser Interferometer Gravitational-wave Observatory (LIGO) has
built three multi-kilometer interferometers, designed to search
for gravitational waves (GWs). One of the possible targets of such a
search is the stochastic background of gravitational waves. 
Many possible sources of such a background have been
proposed (see, e.g., \cite{maggiore,allen,buonanno} for
reviews). Some of these sources are astrophysical in nature, such
as rotating neutron stars, supernovae or low-mass X-ray binaries. Others are
cosmological, such as the amplification of quantum vacuum 
fluctuations during inflation~\cite{par_ampl,star}, phase transitions~\cite{PT}, and 
cosmic strings~\cite{CS}. Most of these sources are expected to
be very weak and below the sensitivity of the LIGO interferometers.
Furthermore, they are constrained by several observations. 

The measurement of the cosmic microwave background by the Cosmic
Background Explorer (COBE) bounds the logarithmic spectrum of
gravitational waves 
\footnote{Note that $\Omega_{\rm GW}$ here should not be 
confused with the ratio of the total energy density stored in 
gravitational waves and the critical 
density of the Universe, i.e., $\Omega_g = \rho_{\rm GW} / \rho_c$.  
The quantity $\Omega_{\rm GW}$ is the so-called GW spectrum
per unit log of frequency: $\Omega_{\rm GW} \equiv (1/\rho_c) \; 
d \rho_{\rm GW} / d \log f$. It would have been more appropriate 
to denote the GW spectrum by $d \Omega_{\rm GW}/d \log f$ and not 
$\Omega_{\rm GW}$.}
to $\Omega_{\rm GW}(f) h_{100}^2 < 8\times 10^{-14}$
at $\sim10^{-16}$ Hz~\cite{cobe1}, where $h_{100} = H_0 / (100 {\rm \;
km / s/ Mpc}) \approx 0.72$ is the "reduced" Hubble parameter \cite{hubble}. 
Since in standard (slow-roll) inflationary models, the spectrum produced 
by the parametric amplification of quantum-vacuum fluctuations~\cite{par_ampl}
is expected to be (almost) flat at higher frequencies~\cite{cobe2}, a similar bound 
applies at higher frequencies, as well. In some inflationary 
models in which a cosmological phase with equation of state stiffer 
than radiation comes before the radiation era, 
the spectrum at high frequency could increase as function of frequency, thus 
avoiding the COBE bound. For example this happens in quintessential 
inflation ~\cite{PV99}. The GW spectrum could mildly increase as 
function of frequency in scenarios in which inflation occurs with 
an equation of state $w  < -1$~\cite{LG} --- some examples 
are given in Ref.~\cite{FF} where inflation is obtained from a non-canonical Lagrangian. 
In other scenarios of superstring cosmology, as the cyclic/ekpyrotic models~\cite{BST}, 
the GW spectrum also increases as function of frequency, but its normalization 
makes it unobservable by ground- and space-based detectors. 

The arrival times of the millisecond pulsars can be used to place 
a bound at $\sim 10^{-8}$ Hz~\cite{pulsar}: 
$\Omega_{\rm GW}(f) h_{100}^2 < 9.3 \times 10^{-8}$.
Doppler tracking of the Cassini spacecraft can be used to arrive
at yet another bound, in the $10^{-6}-10^{-3}$ Hz band
\cite{doppler}: $\Omega_{\rm GW}(f) h^2_{100} < 0.014$. The Big-Bang
Nucleosynthesis (BBN) model and observations can be used to constrain
the integral of the GW spectrum $\int \Omega_{\rm
GW} h_{100}^2 \; d(\ln f) < 6.3 \times 10^{-6}$~\cite{BBN,maggiore,allen}. 
Finally, the ground-based
interferometers and resonant bars can probe the spectrum of
gravitational waves in the band 10 Hz - few kHz. The most recent
bound from LIGO is $\Omega_{\rm GW} h^2_{100} < 4.2 \times 10^{-4}$ for a
flat spectrum in the 69-156 Hz band~\cite{stochpaper}.

In this paper, we focus on the implications of the recent LIGO
result on pre-Big-Bang (PBB) models \cite{pbbrep}, and we
investigate their accessibility to future LIGO searches. The PBB
models predict a stochastic GW 
spectrum whose amplitude can increase as a function of frequency in
some frequency ranges. Hence, they can avoid the bounds due to the
CMB, pulsar timing, and Doppler tracking, and predict relatively
large background in the frequency band where LIGO is sensitive. In
Sec. II we briefly review the GW spectrum in the minimal PBB models. 
In Sec. III we discuss the latest result from the LIGO search for 
the stochastic GW background. In Sec. IV we study how the
new LIGO results, and the expected future results, constrain the
free parameters of the minimal PBB models. In Sec. V we discuss how 
modifications of the minimal PBB model can affect the observability 
of the stochastic GW background. Finally, we conclude in Sec. VI.

\section{The gravitational wave spectrum in the minimal Pre-Big-Bang model}

In the PBB scenario (see, e.g., ~\cite{pbbrep,pbb,pbb2,pbb3}), 
the initial state of the Universe is assumed to be the string
perturbative vacuum, where the Universe can be described by the
low-energy string effective action. The kinetic energy of the 
dilaton field drives the Universe through an inflationary 
evolution (henceforth denoted dilaton inflationary phase), 
which is an accelerated expansion in the string frame, or 
accelerated contraction (gravitational collapse) in the 
(usual) Einstein frame. The spacetime curvature increases in 
the dilaton inflationary phase, eventually reaching the order of 
the string scale. 
At this point, the low-energy string effective action
is no longer an accurate description of the Universe, and higher
order corrections (higher-curvature and/or quantum-loop corrections) 
should be included in the string action. These corrections are expected 
to reduce or stop the growth of the curvature, {\it removing} 
the would-be Big-Bang singularity. 
The exact evolution of the Universe in this high curvature and/or 
strong-coupling phase (henceforth denoted by {\it stringy} phase) is currently not 
known~\cite{pbbrep}. The end of the stringy phase is what one 
could refer to as the ``Big-Bang'' --- the Universe's transition 
into the radiation phase, which is then followed by the matter-dominated 
and acceleration-dominated phases.

Although the transition between the inflationary PBB phase and the
post-Big-Bang phase is not well understood, some models have been
proposed in the literature which can partially describe it. 
In the following, we focus on the model derived in Ref.~\cite{GMV} where, 
in the string frame, the dilaton-inflationary phase is followed by a phase of constant
curvature with the dilaton field growing linearly in time. 
It is then assumed that at the end of this stringy phase the dilaton reaches the
present vacuum expectation value and stops. This model has been 
denoted in the literature as the ``minimal'' PBB model. 
Within this model, the stochastic GW background has been 
evaluated~\cite{BGGV,RB,BMU}. For simplicity, in this paper 
we use the result for the logarithmic spectrum of gravitational waves
~\cite{par_ampl} as evaluated in Ref.~\cite{BMU}:
\begin{eqnarray}
\label{spectrum}
h_{100}^2 \Omega_{\rm GW}(f) & = & b(\mu) \frac{(2 \pi f_s)^4}{H_{100}^2
M_{Pl}^2} \Bigg( \frac{f_1}{f_s} \Bigg)^{2\mu +1} \Bigg(
\frac{f}{f_s} \Bigg)^{5-2\mu}
\nonumber \\
& \times & \Bigg| H_0^{(2)}\Bigg( \frac{\alpha f}{f_s} \Bigg)
J'_{\mu} \Bigg( \frac{f}{f_s} \Bigg) \nonumber \\
& + & H_1^{(2)} \Bigg( \frac{\alpha f}{f_s} \Bigg)
J_{\mu} \Bigg( \frac{f}{f_s} \Bigg) \nonumber \\
& - & \frac{ (1 - \alpha )}{2 \alpha} \frac{f_s}{f} H_0^{(2)} \Bigg(
\frac{ \alpha f}{f_s} \Bigg) J_{\mu} \Bigg( \frac{f}{f_s} \Bigg) \Bigg|^2\, ,
\label{main}
\end{eqnarray}
\begin{eqnarray}
b(\mu) & = & \frac{\alpha}{48} 2^{2\mu} (2\mu - 1)^2 \Gamma^2(\mu)\,,
\nonumber \\
\alpha & = & \frac{1}{1 + \sqrt{3}}\,, \nonumber
\end{eqnarray}
where $H^{(2)}_{0,1}, J_\mu,$ and $\Gamma$ are the Hankel, Bessel and Gamma
functions, respectively, $H_{100} = 100$ km/s/Mpc, and $M_{Pl}$ is
the Planck mass; $f_s$ is the GW frequency redshifted until today 
of fluctuations exiting the Hubble radius at the time of the transition 
between the dilaton and the stringy phase; $\mu$ is a dimensionless free parameter that
measures the growth of the dilaton during the stringy phase, 
effectively determining the slope of the spectrum in the
high-frequency limit (see below). The low-frequency limit of Eq.~(\ref{spectrum}) 
is given by~\cite{BMU}:
\begin{eqnarray}
h_{100}^2 \Omega_{\rm GW}(f) & \simeq & \frac{(2\mu -1)^2}{192 \mu^2
\alpha} \frac{(2\pi f_s)^4}{H_{100}^2 M_{Pl}^2} \Bigg(
\frac{f_1}{f_s} \Bigg)^{2\mu + 1}
\Bigg( \frac{f}{f_s} \Bigg)^3 \nonumber \\
& \times & \Bigg\{ (2 \mu \alpha - 1 + \alpha )^2 \nonumber \\
& + & \frac{4}{\pi^2}
\Bigg[ ( 2 \mu \alpha -1 + \alpha ) \Bigg( \ln \frac{\alpha f}{2 f_s} +
\gamma_E \Bigg) - 2 \Bigg]^2 \Bigg\} \nonumber \\
\gamma_E & = & 0.5772\,,
\label{lowfreq}
\end{eqnarray}
while the high-frequency limit is~\cite{BMU}:
\begin{equation}
h_{100}^2 \Omega_{\rm GW}(f) \simeq \frac{4 b(\mu )}{\pi^2 \alpha}
\frac{(2 \pi f_1)^4 }{H_{100}^2 M_{Pl}^2} \Bigg( \frac{f}{f_1}
\Bigg)^{3-2\mu} \label{highfreq}\,.
\end{equation}
The parameter $f_1$ appearing in the above equations is the 
GW frequency redshifted until today of fluctuations exiting the Hubble 
radius when the stringy phase ends. This is the largest frequency (smallest scale) 
for which fluctuations are amplified --- hence, $f_1$ is also the 
high-frequency cut-off of the GW spectrum.

\begin{figure}[hbtp]
\includegraphics[width=3.3in]{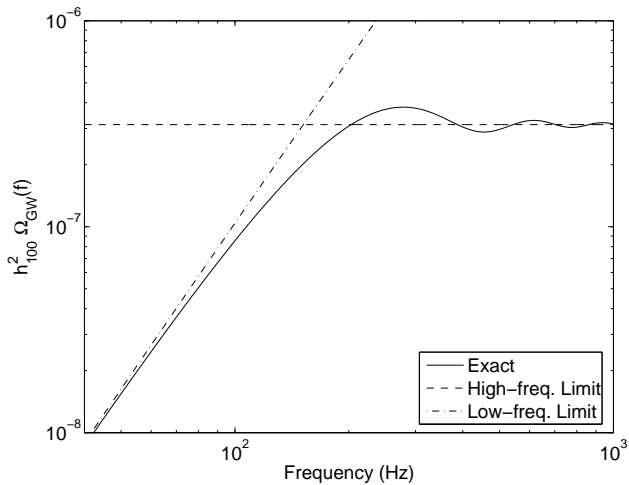}
\caption{$h_{100}^2 \Omega_{\rm GW}(f)$ vs $f$, as predicted by the PBB
model with $f_s=100$ Hz, $f_1 = 4.3\times 10^{10}$ Hz, and
$\mu=1.5$.}
\label{example}
\end{figure}

Thus, the GW spectrum in the minimal PBB scenario increases as $f^3$
for $f \ll f_s$, goes as $f^{3-2\mu}$ for $f_s \ll  f \ll f_1$,
and vanishes exponentially for $f>f_1$. 
An example of such a spectrum is shown in Fig. 3 of Ref.~\cite{BMU}, 
and we reproduce it in Fig.~\ref{example}.

Let us now focus on the free parameters of the model. The
parameter $\mu$ is, by definition, limited to positive values. We
will only consider the case $\mu \le 1.5$ --- for $\mu>1.5$ 
the decreasing spectrum would easily violate 
the existing experimental bounds~\cite{BMU}. The
parameter $f_s$ varies over the range $0 < f_s < f_1$. Since the
spectrum sharply decreases for frequencies below $f_s$, LIGO's
reach for models where $f_s$ is above the LIGO band quickly
diminishes. In particular, to avoid the $f^3$ dependence in the 
LIGO frequency band, $f_s \lesssim 30$ Hz is necessary. Furthermore, 
Eq.~(\ref{highfreq}) shows that in the high-frequency limit
the spectrum does not depend on $f_s$. Hence, if $f_s
\lesssim 30 $ Hz, it does not matter what it is, as far as the
accessibility to LIGO is concerned. Finally, the parameter $f_1$
can be approximated as \cite{BMU}~\footnote{Note that 
in evaluating this equation 
and also the GW spectrum above, we did not include the very recent phase 
of acceleration of the Universe, but limited to radiation and matter eras. 
We expect that if the acceleration era were included, the effect on the 
results presented here would be mild.}:
\begin{equation}
f_1 \simeq 4.3 \times 10^{10} {\rm \; Hz\; } \Bigg( \frac{H_s}{0.15 M_{Pl}}
\Bigg) \Bigg( \frac{t_1}{\lambda_s} \Bigg)^{1/2}\,,
\label{f1approx}
\end{equation}
where $H_s$ is the (constant) Hubble parameter during the stringy phase, $t_1$
is the time when the string phase ends, and $\lambda_s$ is the
string length. The values $H_s \approx 0.15 M_{Pl}$ and $t_1 \approx
\lambda_s$ are the most natural ones~\cite{BMU}~
\footnote{The most natural value for $H_s$ was obtained 
in Ref.~\cite{BMU} by assuming $H_s \sim 1/\lambda_s$ and 
$\lambda_s^2 \sim (2/\alpha_{\rm GUT})\,L_{\rm Pl}^2$ with 
$\alpha_{\rm GUT} \sim 1/20$.}, but they might vary by 
an order of magnitude. Since $\Omega_{\rm GW}(f) \sim f_1^4$ [see Eq.~(\ref{highfreq})], 
this variation leads to a very large variation in the amplitude of the GW
spectrum. Hence, although the theoretically predicted 
value for $f_1$ is more robust than those for 
$f_s$ and $\mu$, we shall explore the possibility of 
varying $f_1$ around its most natural value~\footnote{We 
note that in the more common version of the minimal PBB model~\cite{pbbrep,BGGV,BMUV}, 
the frequency $f_1$ is obtained by imposing that the energy density becomes critical at the 
beginning of the radiation phase and that the photons we observe today 
originated from the amplified vacuum fluctuations during the dilaton-driven 
inflationary phase. Within these assumptions Eq.~(\ref{f1approx}) can be re-written as 
$f_1 \simeq g_1^{1/2}\,(H_s/(0.15\,M_{\rm Pl}))^{1/2}\,
(H_0\,M_{\rm Pl})^{1/2}\,\Omega_{\gamma}^{1/4}$, where $\Omega_\gamma = 
4 \times 10^{-5}\, h^{-2}_{100}$ and $g_1$ is the string coupling at 
the end of the stringy phase.}. 

\section{Searching for stochastic gravitational waves with LIGO}

The method of searching for stochastic gravitational waves with
interferometers has been studied by many authors~\cite{christensen,flanegan,allenromano}. 
Following Allen and Romano~\cite{allenromano}, we can define the following 
cross-correlation estimator:
\begin{eqnarray}
Y & = & \int_{-\infty}^{+\infty} Y(f) \; df\,, \\
& = & \int_{-\infty }^{+\infty } df \int_{-\infty }^{+\infty } df' \;
\delta_T (f-f') \; \tilde{s}_1(f)^{*} \; \tilde{s}_2(f') \; \tilde{Q}(f')\,,
\nonumber
\label{ptest}
\end{eqnarray}
where $\delta_T$ is a finite-time approximation to the Delta function,
$\tilde{s}_1$ and $\tilde{s}_2$ are the Fourier transforms of the
strain time-series of two interferometers, and $\tilde{Q}$ is
the optimal filter. Assuming that the detector noise
is Gaussian, stationary, uncorrelated between the two interferometers,
and uncorrelated with and much larger than the GW signal, the
variance of the estimator $Y$ is given by:
\begin{equation}
\sigma_Y^2 \approx \frac{T}{2} \int_0^{+\infty} df P_1(f) P_2(f)
\mid \tilde{Q}(f) \mid^2\,,
\label{sigma}
\end{equation}
where $P_i(f)$ are the power spectral densities of the two interferometers,
and $T$ is the measurement time. Finally, it can be shown that the
optimal filter can be written in the form \cite{allenromano}:
\begin{equation}
\tilde{Q}(f) = N \frac{\gamma(f) \Omega_t(f)}{f^3 P_1(f) P_2(f)}\,,
\end{equation}
where $\gamma(f)$ is the overlap reduction function (arising from the
different locations and orientations of the two interferometers), and
$\Omega_t(f)$ is the template spectrum to be searched. Assuming the
template spectrum $\Omega_t(f) = \Omega_{\alpha}
(f/100 {\rm \; Hz})^{\alpha}$, the
normalization constant $N$ can be chosen such that $<Y> = \Omega_{\alpha} T$.

\begin{figure}[hbtp]
\includegraphics[width=3.3in]{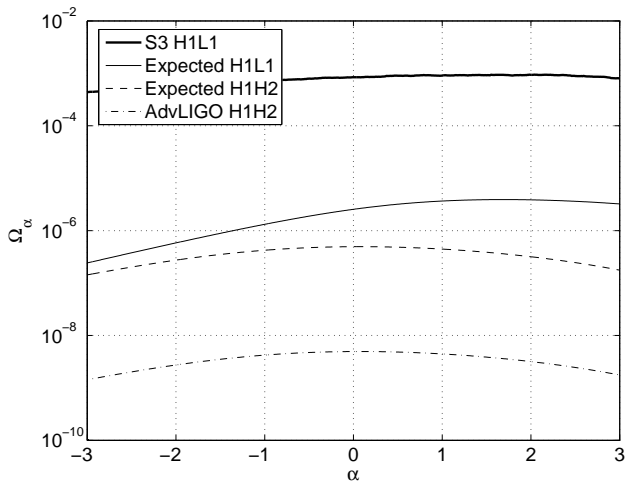}
\caption{The 90\% UL on $\Omega_{\alpha}$ is shown as a function of the 
spectral slope $\alpha$ for the most recent LIGO result. Expected
sensitivities of LIGO and of Advanced LIGO are also shown.}
\label{expreach}
\end{figure}

This analysis procedure was implemented in the recent analysis of
the LIGO data, using the 4 km interferometers at Hanford, WA and
Livingston, LA, for the science run S3 \cite{stochpaper}. This
analysis yielded the 90\% upper limit of $\Omega_0 < 8.4 \times
10^{-4}$ for the flat template spectrum $\Omega_t(f) = \Omega_0$.
Once $Y(f)$ is estimated for the flat spectrum, one can apply
simple scaling by the appropriate power law to obtain the
estimates for different values of $\alpha$ (similar procedure can be 
followed for an arbitrary spectral shape). Fig. \ref{expreach}
shows the 90\% UL on $\Omega_{\alpha}$ as a function of the
spectral slope $\alpha$ for the S3 run, as well as the expected
reach for LIGO and for Advanced LIGO. Here and in the following 
by expected LIGO (H1L1 and H1H2) we mean LIGO design sensitivity and 
one year of observation, and by Advanced LIGO we assume 
a sensitivity 10 times better than the LIGO design and one year of observation.
[LIGO has started the year-long run at design sensitivity in November 2005.]

\section{Scanning the parameter space}

We now study the accessibility of the minimal PBB model 
discussed in Sec. II to the most recent and future runs 
of LIGO, and to Advanced LIGO.
Previous investigations, which did not use real data,  
were done in Refs.~\cite{allenbrustein,UV}.

\begin{figure}[hbtp]
\includegraphics[width=3.3in]{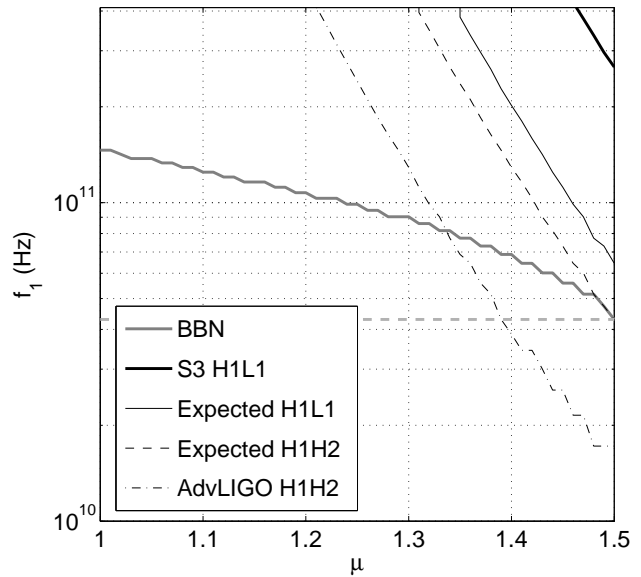}
\caption{The 90\% UL exclusion curves are shown in the $f_1-\mu$
plane for $f_s=30$ Hz (the excluded regions are above the
corresponding curves). We show the latest result from LIGO, and
the future expected reach of LIGO and of Advanced LIGO. The
limit from the BBN is also shown. The horizontal gray dashed line
denotes the most natural value of $f_1$, given by Eq. (\ref{f1approx}).}
\label{f1mu}
\end{figure}

As discussed in Section 2, the amplitude of the GW spectrum in the PBB models
is proportional to $f^3$ at frequencies below $f_s$. Hence, the sensitivity
of LIGO to PBB models decreases as $f_s$ is increased. 
To avoid the $f^3$ dependence of the spectrum in the LIGO frequency band,
we choose $f_s=30$ Hz.
For such choice of $f_s$, the LIGO band falls in the relatively flat part
of the GW spectrum. We vary $f_1$ by a factor of 10 around the most 
natural value estimated in Eq.~(\ref{f1approx}) (i.e., between 
$4.3 \times 10^9$ and $4.3 \times 10^{11}$)
and we vary $\mu$ between $1$ and $1.5$ 
(models with $\mu<1$ are out of reach of LIGO, as shown below). 
For each point in the $\mu-f_1$ plane,
we evaluate $\Omega_{\alpha} = \Omega_{\rm GW}(f = 100 {\rm \; Hz})$
predicted by the model, and we check whether it is excluded by the
experimental (or future expected) results. We also integrate the
predicted spectrum and check whether it passes the BBN bound~\cite{BBN,maggiore,allen}:
\begin{equation}
\int \Omega_{\rm GW}(f) h_{100}^2 d(\ln f) < 6.3 \times 10^{-6}\,,
\label{bbnlimit}
\end{equation}
assuming the number of neutrino species $N_{\nu}<3.9$ \cite{copi,BMU}.
We use $h_{100}=0.72$ as the reduced Hubble parameter \cite{hubble}. Fig.
\ref{f1mu} shows the 90\% UL exclusion curves obtained in this way. The
latest result from LIGO (S3 run) is just beginning to probe 
this parameter space. The future runs of LIGO (and of Advanced LIGO) are
expected to probe a more significant part of the parameter space, becoming 
comparable to or even surpassing the BBN bound. 
As expected, LIGO is most sensitive to models with $\mu=1.5$, which
corresponds to the flat spectrum at high frequencies.
As $\mu$ decreases from 1.5, the spectral slope increases, 
and the spectrum in the LIGO band drops quickly
below LIGO sensitivity. Although the BBN bound also weakens
for $\mu<1.5$, the effect is not as dramatic because this bound is
placed on the integral of the spectrum over a large frequency range.
Note that the LIGO S3 run is sensitive to PBB models with 
$f_1 \ge 2.7 \times 10^{11}$ Hz, relatively large compared to the most natural
value estimated in Eq. (\ref{f1approx}).
This is true independent of $f_s$: for $f_s < 30$ Hz, 
Fig. \ref{f1mu} would not change, while for $f_s > 30$ Hz all bounds
would weaken. Finally, the Advanced LIGO is expected to reach models with 
the most natural value of $f_1 = 4.3 \times 10^{10}$ Hz.

\begin{figure}[hbtp]
\includegraphics[width=3.3in]{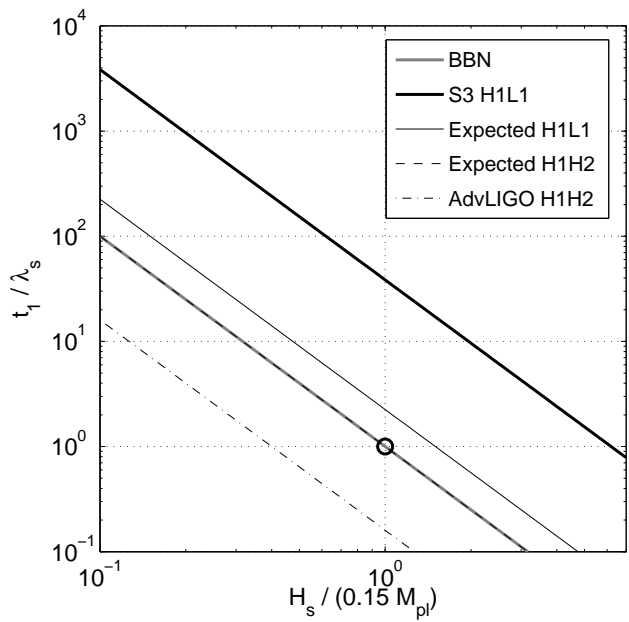}
\caption{The 90\% UL exclusion curves are shown in the
$t_1/\lambda_s$ vs $H_s/(0.15 M_{Pl})$ plane, for $\mu=1.5$ and $f_s=30$
Hz (the excluded regions are above the corresponding curves). We
show the latest result from LIGO, and the future expected reach
of LIGO and of Advanced LIGO. The limit from the BBN is also shown.
The black circle denotes the most natural point, as given in Eq. 
(\ref{f1approx}).}
\label{t1Hs}
\end{figure}

It is also possible to use Eq.~(\ref{f1approx}) to turn a bound
on $f_1$ into an exclusion curve in the $t_1/\lambda_s$ vs
$H_s/(0.15 M_{Pl})$ plane. In this way, the GW experiments can be 
used to constrain string-related parameters in the framework of the PBB
model. As an example, we choose $\mu = 1.5$ and $f_s = 30$ Hz 
as the optimal case for LIGO, and
determine the 90\% UL exclusion curves for different experiments.
These curves are shown in Fig.~\ref{t1Hs}. Again, the latest
LIGO result is weaker than the BBN bound, but the future LIGO and
Advanced LIGO searches are expected to explore a larger, more physical
part of this parameter space. 

\begin{figure}[hbtp]
\includegraphics[width=3.3in]{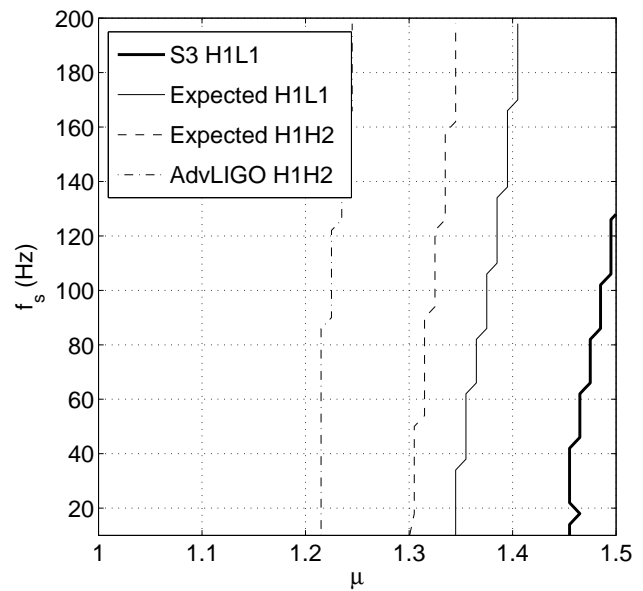}
\caption{The 90\% UL exclusion curves are shown in the
$f_s-\mu$ plane, for $f_1 = 4.3 \times 10^{11}$ Hz
(the excluded regions are to the right from the corresponding curves). We
show the latest result from LIGO, and the future expected reach
of LIGO and of Advanced LIGO. The indirect limit from the BBN excludes the
whole region shown in this plane.}
\label{fsmu}
\end{figure}

One can also examine the accessibility of the models in the $f_s - \mu$ plane.
For the relatively large value $f_1 = 4.3 \times 10^{11}$ Hz, 
which makes the model's stochastic GW background 
accessible to the LIGO S3 run, we performed a scan in the $f_s - \mu$ plane.
Fig.~\ref{fsmu} shows the results. Note that for the flat spectrum ($\mu=1.5$),
the S3 run of LIGO is sensitive to models with $f_s \lesssim 120$ Hz; 
future runs of LIGO
and Advanced LIGO are expected to probe higher values of $f_s$ as well. Also
note that the exclusion curves in Fig. \ref{fsmu} are almost vertical (i.e. 
not very sensitive to $f_s$). This is a consequence of the large value of 
$f_1$ - for smaller values of $f_1$, the accessibility of models to LIGO
would depend more strongly on the value of $f_s$. 

\begin{figure}[hbtp]
\includegraphics[width=3.3in]{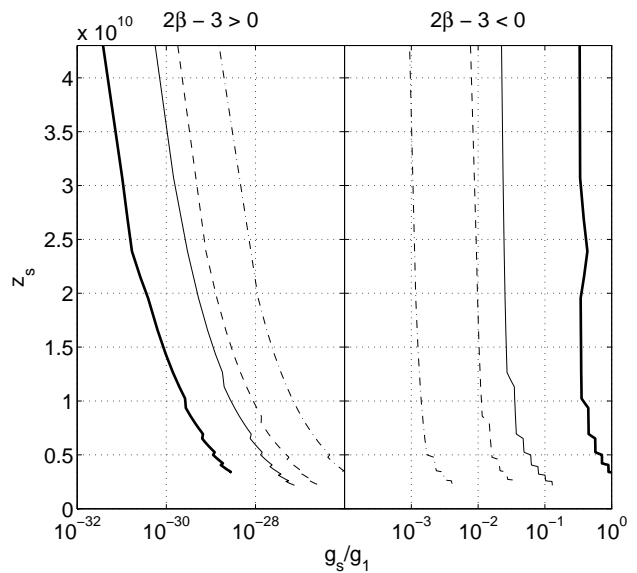}
\caption{The 90\% UL exclusion curves are shown in the
$z_s - g_s$ plane, for $f_1 = 4.3 \times 10^{11}$ Hz. We
show the latest result from LIGO (thick solid), and the future expected reach
of LIGO (thin solid for the H1L1 pair, dashed for the H1H2 pair) and 
of Advanced LIGO (dash-dotted). The two sets of curves correspond to 
positive (left) and negative (right) signs of $(2\beta - 3)$.}
\label{zsgs}
\end{figure}

Several papers in the literature~\cite{BGGV,RB,peak}, parametrize the 
GW spectrum in the minimal PBB model in terms of $z_s = f_1/f_s$ and 
$g_s$, defined by $g_s/g_1 = (f_s/f_1)^\beta$, with $\beta$ 
given by $2 \mu = |2 \beta -3|$. The parameter $z_s$ is the total 
redshift during the stringy phase, thus it quantifies its duration, 
while $g_1$ and  $g_s$ are the string couplings at the end and at the 
beginning of the stringy phase, respectively.  
Fig.~\ref{zsgs} shows the curves from 
Fig. \ref{fsmu} converted into the $z_s - g_s$ plane, 
using $f_1 = 4.3 \times 10^{11}$ Hz, and setting $g_1$ 
to its most natural value given by $g_1^2/(4 \pi) = \alpha_{\rm GUT}$.

\section{Going Beyond the Minimal Pre-Big-Bang Model}

In this section we investigate how extensions of the minimal PBB model 
or variations of it can impact the accessibility of the stochastic GW 
background to LIGO and to Advanced LIGO.

The GW spectrum in the minimal PBB model was originally 
evaluated~\cite{BGGV,RB,BMU} neglecting the higher-curvature 
corrections in the equation of tensorial fluctuations during 
the stringy phase. Gasperini~\cite{MG} evaluated the 
higher-order equation for tensorial fluctuations and showed that 
these corrections modify the amplitude of the perturbation {\it only} 
by a factor of order one. Hence, these corrections are not expected to
affect our results significantly.

\begin{figure}[hbtp]
\includegraphics[width=3.3in]{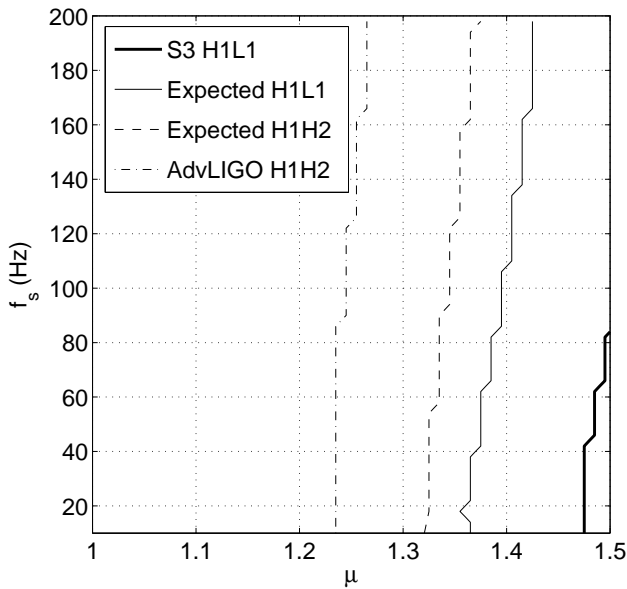}
\caption{The 90\% UL exclusion curves are shown in the $f_s-\mu$
plane for $f_1=4.3\times 10^{11}$ Hz and for $\delta s = 0.5$ (the excluded
regions are to the right of the corresponding curves). We show the latest
result from LIGO, and the future expected reach of LIGO and of
Advanced LIGO. The indirect BBN limit excludes all models shown in this plane.}
\label{fsmuds}
\end{figure}

In Refs.~\cite{gasperini,peak} the authors have examined the effect of 
radiation production via some reheating process occuring below the string
scale. Such a process may be needed to dilute several relic
particles produced during (or at the very end of) the PBB phase, 
whose abundance could spoil the BBN predictions~\cite{BLO}. 
Depending on when and for how long the entropy production occurs, 
it can change both the shape and the amplitude of the GW spectrum 
in the frequency region around $100$ Hz. In general, 
the amplitude of the spectrum at these frequencies is
reduced. If we assume that the reheating process occurs at the end
of the stringy phase (i.e., all of the entropy is produced at
the end of the stringy phase), then the effect of the process is a
simple scaling of the original spectrum by the factor $(1-\delta
s)^{4/3}$, where $\delta s$ is the fraction of the present thermal
entropy density that was produced in the process. Fig.~\ref{fsmuds} 
shows the exclusion curves in the $f_s-\mu$ plane for
$\delta s=0.5$. By comparing to Fig.~\ref{fsmu}, we can see that 
the effect weakens all bounds.

\begin{figure}[hbtp]
\includegraphics[width=3.3in]{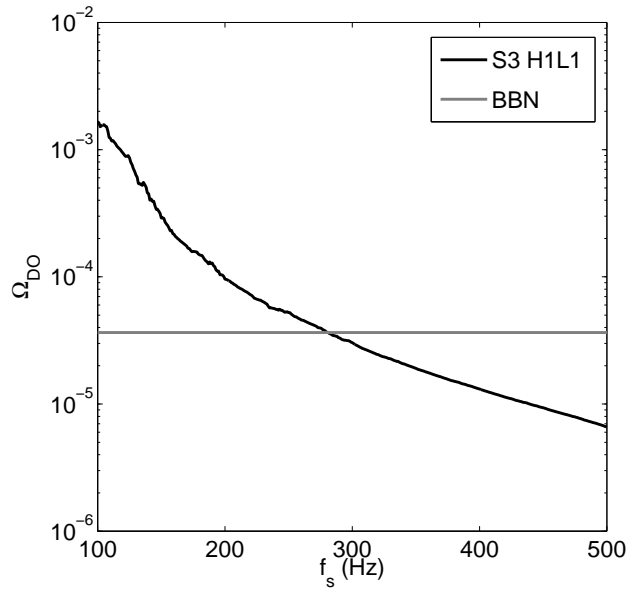}
\caption{The 90\% UL on $\Omega_{DO}$ is shown as a function of $f_s$ 
for the models where
stochastic gravitational background is not produced during the string phase.
The latest LIGO result and the BBN bound are shown.}
\label{AB}
\end{figure}

Another possible, but somewhat arbitrary variation of the model, 
was examined by Allen and Brustein~\cite{allenbrustein}. They assumed 
that stochastic gravitational waves are not produced during the stringy 
phase, but only during the dilaton phase. This is achieved by setting $f_1 = f_s$ and 
assuming that $\Omega_{\rm GW}$ vanishes for $f_s < f$. 
Such a model is not well motivated in the PBB scenario, but it
is phenomenologically interesting as it represents a class of models whose spectrum peaks in
the LIGO band. The spectrum of this model can, therefore, be approximated by:
\begin{equation}
\Omega_{\rm GW}(f) = \left\{
\begin{array}{ll}
\Omega_{\rm DO} \big(\frac{f}{f_s}\big)^3 & f<f_s \,,\\
0 & f>f_s\,.
\end{array}
\right.
\end{equation}
The BBN bound becomes weaker because the integral in Eq.~(\ref{bbnlimit}) 
is performed over a much smaller frequency range, and it can be
written as $\Omega_{\rm DO} < 3.8 \times 10^{-5}$.
Fig.~ \ref{AB} shows the bound from the latest LIGO result as a function
of $f_s$. Note that this bound is already better than the BBN bound for
$f \gtrsim 300$ Hz.

Finally, as first noticed in Ref.~\cite{gasperini}, it is well possible that 
many more cosmological phases are present between the pre- and the 
post-Big-Bang 
eras --- some examples are given in Refs.~\cite{gasperini,BMUV}. 
If this is the case, the GW spectra 
during the high-curvature and/or strong coupling region 
will be characterized by several branches with increasing and decreasing 
slopes. Due to the dependence of the spectra on a larger number of parameters, 
it would be more difficult to constrain these non-minimal scenarios, even when 
LIGO overcomes the BBN bound. 

\section{Conclusions}

Using the most recent LIGO search for the stochastic
gravitational background~\cite{stochpaper}, 
we determined the exclusion regions in 
the parameter space of the PBB minimal model~\cite{pbbrep}. 
We found that the most recent S3 run can access 
the stochastic GW background only if $f_1$ is 
larger than the most natural value $4.3 \times 10^{10} {\rm Hz}$ 
(i.e. only if $f_1 > 2.7 \times 10^{11}$ Hz; see Fig. \ref{f1mu}). 
In this case the S3 run of LIGO can exclude the 
region in the $(f_s,\mu)$ parameter space with $\mu \approx 1.5$ (i.e., almost 
a flat spectrum) and $f_s \lesssim 120\, {\rm Hz}$ (see Fig. \ref{fsmu}). 
A one-year run of LIGO at the design sensitivity will be able to start excluding 
regions with sligthly increasing GW spectrum (as a function of frequency), 
while Advanced LIGO could exclude 
spectra with slopes of at most $\approx 0.5$. Models with larger values of 
$f_s$ will
become accessible, as well as with lower values of $f_1$, including the
most natural value $f_1 = 4.3 \times 10^{10}$ Hz.

As shown in Fig. \ref{f1mu}, the BBN bound already {\it excludes} all models 
accessible to the LIGO S3 run. However, it should be noted that: (i) the LIGO
S3 bound is a result of a direct measurement of the stochastic background of 
gravitational waves while BBN bound is not, and (ii) future searches 
by LIGO and by Advanced LIGO are expected to approach and even surpass 
(in some parts of the parameter space) the BBN bound.

Analysis of the 
search in the parameter space more commonly used in the literature 
(see Fig. \ref{zsgs}) shows that LIGO and Advanced LIGO can bound the 
duration of the stringy phase and the string coupling at the beginning 
(end) of the stringy phase (dilaton inflationary phase). Similarly, 
as shown in Fig. \ref{t1Hs}, by constraining $f_1$ these experiments can 
constrain other string-related parameters, such as $H_s$ (the Hubble parameter 
during the stringy phase) and $t_1/\lambda_s$ (the ratio of the end-time 
of the stringy phase and of the string length) or the value of the string 
coupling at the end of the stringy phase $g_1$.

As emphasized above, the stringy phase 
is not well understood, yet. Many variations 
to the minimal PBB model analyzed in this paper are possible and have been
proposed~\cite{peak,gasperini,BMUV}. They can significantly change the 
shape and the amplitude of the spectrum in the frequency range around $100$ Hz, 
hence improving or reducing the accessibility of the PBB models to LIGO.  
The presence of multi cosmological phases~\cite{gasperini} during 
the stringy phase will make much harder the determination of the exclusion regions in the 
PBB parameter space. More robust predictions for the stringy phase would be 
strongly desirable.

\begin{acknowledgments}
The authors thank Albert Lazzarini, Joe Romano, and
Stan Whitcomb for many useful discussions, Maurizio Gasperini 
for useful comments, and the LIGO Scientific Collaboration
for making this study possible. V.M.'s  work was 
supported by the NSF Cooperative Agreement No. PHY-0107417.
\end{acknowledgments}


\begin{thebibliography}{999}
%
\bibitem{maggiore}
M. Maggiore, \Journal{\PRep}{331}{283}{2000}.
%
\bibitem{allen}
B. Allen, {\it Lectures at Les Houches School}, [gr-qc/9604033].
%
\bibitem{buonanno}
A. Buonanno, TASI {\it Lectures on Gravitational Waves from the Early Universe}, [gr-qc/0303085].
%
\bibitem{par_ampl} 
L.P. Grishchuk, Sov. Phys. JETP {\bf 40}, 409 (1975); Class. Quantum Grav. 
{\bf 10} 2449 (1993); Class. Quantum Grav. {\bf 14}, 1445 (1997).
%
\bibitem{star} A.A. Starobinsky, Pis'ma Zh. Eksp. Teor. Fiz. {\bf 30}, 719 (1979).
%
\bibitem{PT} A. Kosowsky, M.S. Turner and R. Watkins, 
\Journal{\PRD}{45}{4514} {1992}; \Journal{\PRL}{69}{2026}{1992}; 
A. Kosowsky and M.S. Turner, \Journal{\PRD}{47}{4372}{1993}; 
M. Kamionkowski, A. Kosowsky and M.S. Turner, \Journal{\PRD}{49}{2837}{1994}; 
R. Apreda, M. Maggiore, A. Nicolis and 
A. Riotto, \Journal{\NPB}{631}{342}{2002}.
%
\bibitem{CS} R.R. Caldwell and B. Allen, \Journal{\PRD}{45}{3447}{1992};
R.R. Caldwell, R.A. Battye and E.P.S. Shellard, \Journal{\PRD}{54}{7146}{1996};
T. Damour and A. Vilenkin, \Journal{\PRL}{85}{3761}{2000};
\Journal{\PRD}{64}{064008}{2001}.
%
\bibitem{cobe1}
B. Allen and S. Koranda, \Journal{\PRD}{50}{3713}{1994}.
%
\bibitem{hubble}
C.L. Bennet \etal, \Journal{\APJS}{148}{1}{2003}.
%
\bibitem{cobe2} M. S. Turner, \Journal{\PRD}{55}{R435}{1997}; 
T.L. Smith, M. Kamionkowsky and 
A. Cooray, {\it Direct detection of the inflationary gravitational-wave 
background}, 
[astro-ph/0506422]; L.A. Boyle, P. Steinhardt and N. Turok, 
{\it Inflationary predictions reconsidered}, [astro-ph/0507455].
%
\bibitem{PV99} 
P.J.E. Peebles and A. Vilenkin, \Journal{\PRD}{59}{063505}{1999}; 
M. Giovannini, \Journal{\PRD}{60}{123511}{1999}.
%
\bibitem{LG} L. Grishchuk, {\it Relic gravitational waves and cosmology}, 
[astro-ph/0504018].
%
\bibitem{FF} M. Baldi, F. Finelli and S. Matarrese, 
\Journal{\PRD}{72}{083504}{2005}.
%
\bibitem{BST} L. A. Boyle, P. Steinhardt and N. Turok, 
\Journal{\PRD}{69}{127302}{2004}. 
%
\bibitem{pulsar}
M.P. McHugh \etal, \Journal{\PRD}{54}{5993}{1996}.
%
\bibitem{doppler}
J.W. Armstrong \etal, \Journal{\APJ}{599}{806}{2003}.
%
\bibitem{BBN}
E. Kolb and R. Turner, {\it The Early Universe} (Addison-Wesley, Reading,
MA, 1990).
%
\bibitem{stochpaper}
B. Abbot \etal, 
{\it Upper limit on a stochastic background of gravitational waves}, 
[astro-ph/0507254].
%
\bibitem{pbbrep}
M. Gasperini and G. Veneziano, \Journal{\PRep}{373}{1}{2003}.
%
\bibitem{pbb}
M. Gasperini and G. Veneziano, \Journal{\APP}{1}{317}{1993}.
%
\bibitem{pbb2}
M. Gasperini and G. Veneziano, \Journal{\MPLA}{8}{3701}{1993}.
%
\bibitem{pbb3}
M. Gasperini and G. Veneziano, \Journal{\PRD}{50}{2519}{1994}.
%
\bibitem{GMV}
M. Gasperini, M. Maggiore, and G. Veneziano, \Journal{\NPB}{494}{315}{1997}. 
%
\bibitem{BGGV}
R. Brustein, M. Gasperini, M. Giovannini, and G. Veneziano, \Journal{\PLB}{361}{45}{1995}.
%
\bibitem{RB}
R. Brustein, {\it Spectrum of cosmic gravitational wave background}, 
[hep-th/9604159].
%
\bibitem{BMU}
A. Buonanno, M. Maggiore, C. Ungarelli, \Journal{\PRD}{55}{3330}{1997}.
%
\bibitem{christensen}
N. Christensen, \Journal{\PRD}{46}{5250}{1992}.
%
\bibitem{flanegan}
E. Flanagan, \Journal{\PRD}{48}{2389}{1993}.
%
\bibitem{allenromano}
B. Allen and J. Romano, \Journal{\PRD}{59}{102001}{1999}.
%
\bibitem{allenbrustein}
B. Allen and R. Brustein, \Journal{\PRD}{55}{3260}{1997}.
%
\bibitem{UV}
C. Ungarelli and A. Vecchio, {\it Are pre-big bang models falsifiable 
by gravitational-wave experiment?}, [gr-c/9911104].
%
\bibitem{copi}
C. Copi \etal, \Journal{\PRL}{75}{3981}{1995}.
%
\bibitem{peak}
R. Brustein, M. Gasperini and G. Veneziano, \Journal{\PRD}{55}{3882}{1997}.
%
\bibitem{MG} M. Gasperini, \Journal{\PRD}{56}{4815}{1997}. 
%
\bibitem{gasperini}
M. Gasperini, {\it Relic gravitons from the pre-big bang: what we know 
and what we do not know}, [hep-th/9607146].
%
\bibitem{BLO} A. Buonanno, M. Lemoine, and K.A. Olive,
\Journal{\PRD}{62}{083513}{2000}.
%
\bibitem{BMUV} A. Buonanno, K. Meissner, C. Ungarelli and G. Veneziano, 
JHEP {\bf 01}, 004 (1997).
%
\end{thebibliography}

\end{document}